\documentclass[a4paper]{jpconf}
\usepackage{graphicx}
\usepackage{dcolumn}
\usepackage{subfigure}
\begin{document}
\title{Superconductivity in Y$_3$Ru$_4$Ge$_{13}$ and Lu$_3$Os$_4$Ge$_{13}$: A comparative study}
\author{Om Prakash, A. Thamizhavel, S. Ramakrishnan}
\address{Department of Condensed Matter Physics and Material Science,
Tata Institute of Fundamental Research,
Mumbai-400005, India}
\ead{op1111shukla@gmail.com}
\begin{abstract}
A variety of unconventional superconductors have low carrier density as a common factor. However, the underlying mechanism of superconductivity in such low carrier density systems is not well understood. Besides, small carrier density is an unfavourable component for conventional superconductivity as described by the Bardeen-Cooper-Schrieffer (BCS) theory. Therefore, studying low carrier density systems can lead to a better understanding in such systems. In this paper, we report superconductivity property studies in low carrier density systems, Y$_3$Ru$_4$Ge$_{13}$ and Lu$_3$Os$_4$Ge$_{13}$, using various experimental techniques. Single crystals of Y$_3$Ru$_4$Ge$_{13}$ and Lu$_3$Os$_4$Ge$_{13}$ have been grown using the Czochralski crystal pulling method in a tetra-arc furnace. The x-ray diffraction experiment reveals that both compounds crystallize in cubic structure {(space group $\it{Pm3n}$, no. 223)}. The transport, magnetization and heat capacity measurements show that Y$_3$Ru$_4$Ge$_{13}$ single crystal undergoes a superconducting transition at 2.85~K, whereas, Lu$_3$Os$_4$Ge$_{13}$ becomes superconductor at 3.1~K.
\end{abstract}
\section{Introduction}
A variety of unconventional superconductors present low density of the charge carriers as a common factor, implying that it could be the basis for
a unifying picture to understand the superconductivity in such exotic systems. Low density of charge carriers is one of the characteristic features which is shared by
cuprates, fullerenes and MgB$_2$ \cite{Fleming1991,Holczer1991,Nagamatsu2001}. This is quite surprising since low carrier density is an unfavourable element for superconductivity within the conventional framework of BCS \cite{Bardeen1957} or Migdal${-}$Eliashberg \cite{Migdal1958,Eliashberg1960} theories. Moreover, a small superfluid density, is unavoidably related to poor screening and strong electronic correlations, ingredients which are expected to be also detrimental for conventional superconductivity. On these grounds it is hard to understand why these low carrier materials are the best superconductors. As far as the superconductivity exhibited by inter-metallic compounds is concerned, the role of electron-phonon interaction cannot be overlooked. However, one may have to look beyond the conventional framework of BCS or Migdal${-}$Eliashberg theories in order to understand the unconventional superconductivity in these compounds. From the experimental side, it is important to look for new superconducting materials with low carrier density. 
\section{Experimental Details}
Single crystals of Y$_3$Ru$_4$Ge$_{13}$ and Lu$_3$Os$_4$Ge$_{13}$ have been grown using Czochralski crystal pulling method in a tetra-arc furnace under high purity argon atmosphere. Stoichiometric ammount of Y$_3$Ru$_4$Ge$_{13}$ and Lu$_3$Os$_4$Ge$_{13}$ (10~g each) was taken and melted 4-5 times in the tetra-arc furnace to make a homogeneous polycrystalline mixture. Single crystals were pulled using a tungsten seed rod at the rate of 10 mm/h for about 6 h to get 5-6~mm long and 3-4~mm thick crystals. The phase purity was characterized by powder X-ray diffraction using PANanalytical X-ray diffractometer. Single crystals were oriented along the crystallographic direction [100] using Laue back reflection using Huber Laue diffractometer and cut to desired shape and dimensions using a spark erosion cutting machine. Resistivity measurements were done in a home made setup using standard four-probe technique. Magnetization measurements were done in commercial SQUID magnetometer (MPMS5, Quantum Design, USA) and heat capacity measurements were done using PPMS.
\section{Results and Discussion}
\begin{figure}[h]
\begin{minipage}[t]{0.45\linewidth}
\includegraphics[height=6cm]{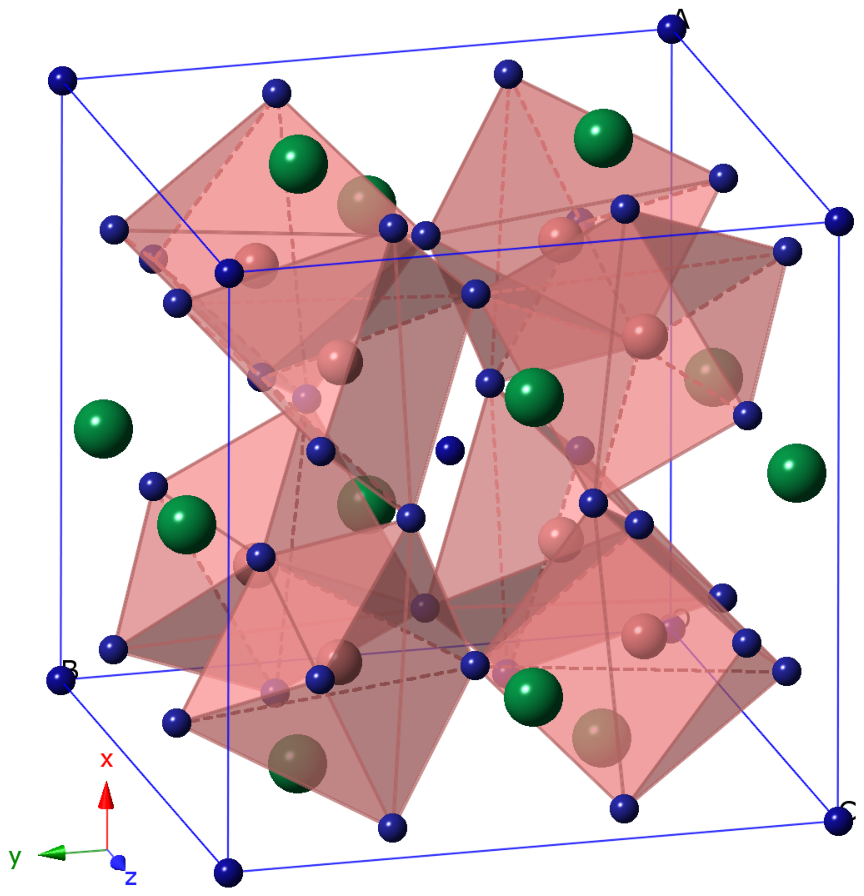}
\caption {Crystal structure of Y$_3$Ru$_4$Ge$_{13}$. Yttrium atoms are shown in green in 6d position, Ruthenium are shown in light pink in 8e position and Germanium are shown in
dark blue in 2a and 24k Wyckoff positions.}
\label{fig:fig1}
\end{minipage}
\hfill
\begin{minipage}[t]{0.45\linewidth}
\includegraphics[height=6cm]{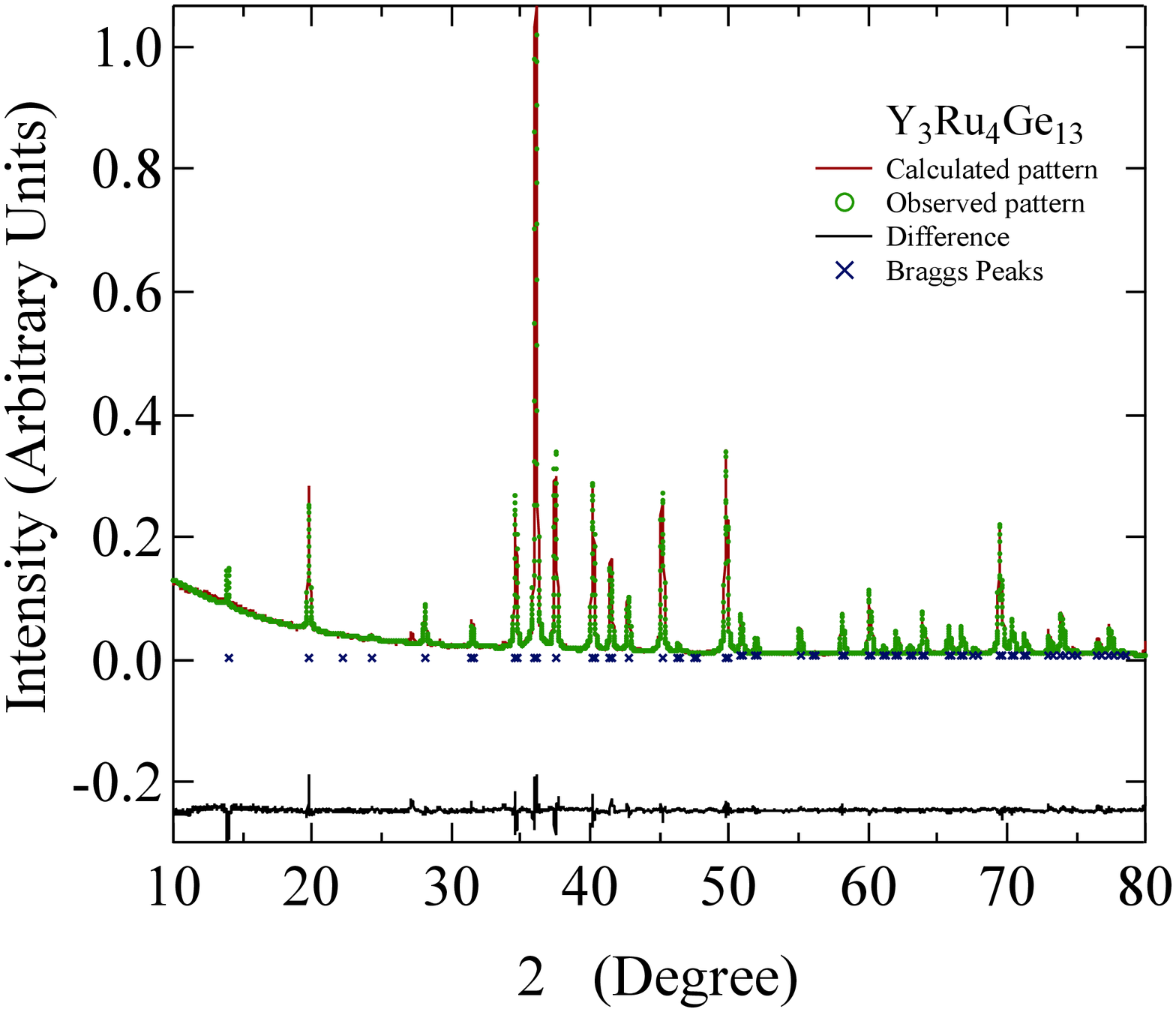}
\caption{Rietveld analysis of the powder XRD pattern of Y$_3$Ru$_4$Ge$_{13}$. No impurity peaks are observed indicating single phase nature of the compound. Similar PXRD pattern is also observed for Lu$_3$Os$_4$Ge$_{13}$.}
\label{fig:fig2}
\end{minipage}
\end{figure}

\begin{figure}[ht]
\hfill
\subfigure{\includegraphics[width=7cm]{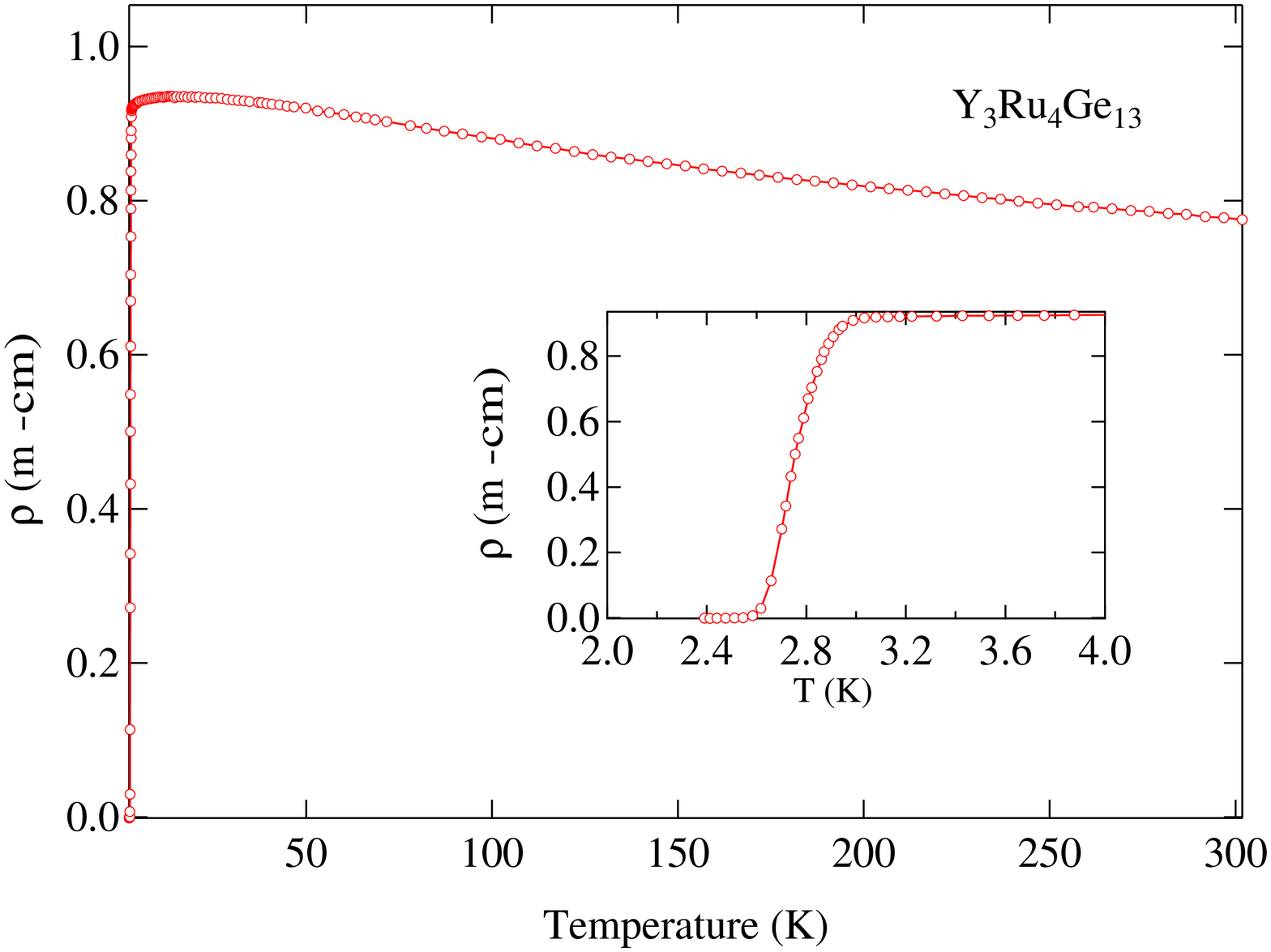}}
\label{fig:fig2a}
\hfill
\subfigure{\includegraphics[width=7cm]{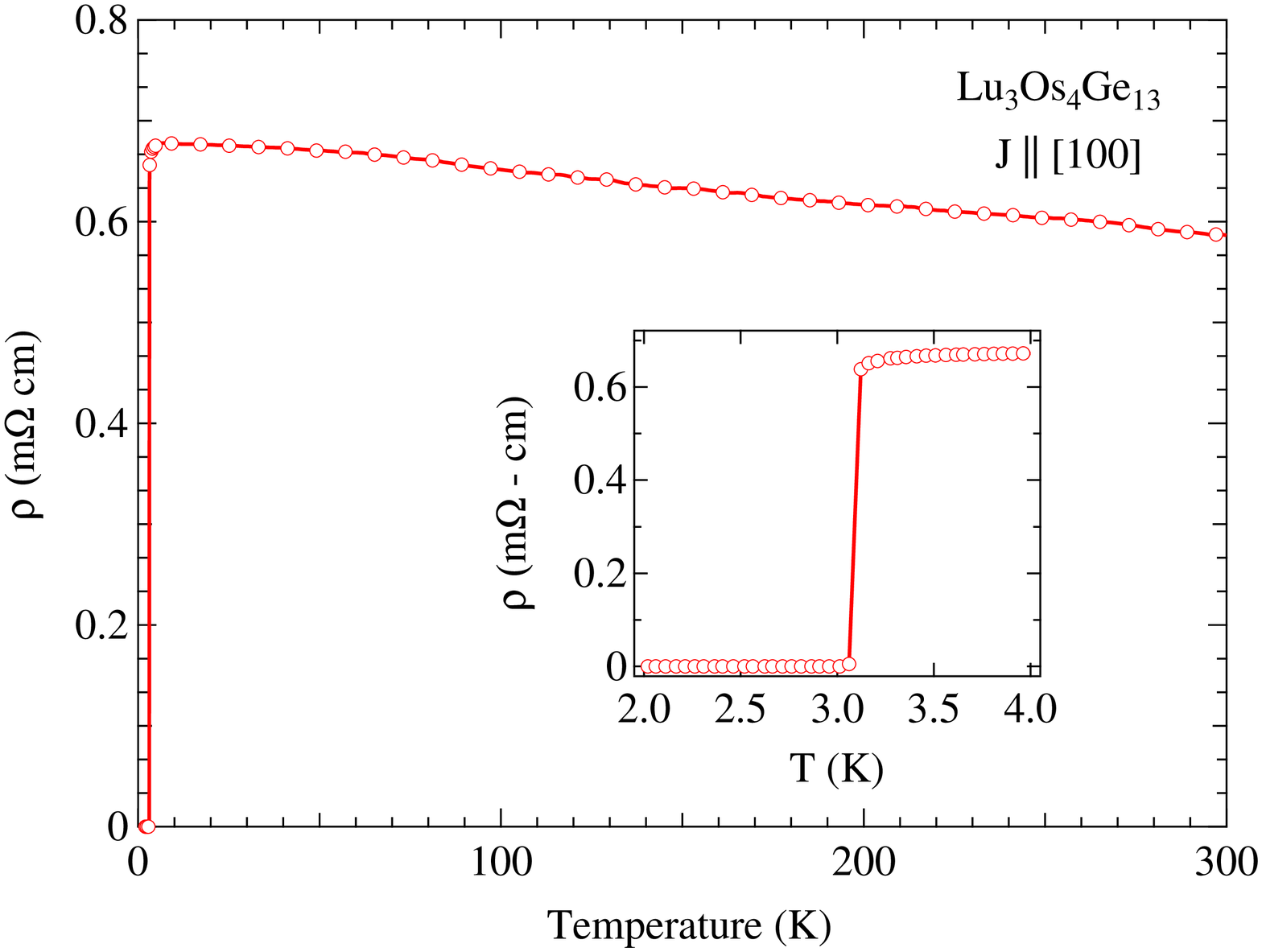}}
\label{fig:fig1b}
\hfill
\caption{The temperature dependence of the electrical resistivity ($\rho$) along the (100) directions of cubic (Pm3n) Y$3$Ru$4$Ge$13$ and Lu$_3$Os$_4$Ge$_{13}$. Insets show the low temperature data indicating superconducting transition in both compounds. Resistivity data from 2 to 300K clearly shows the semi-metallic nature of both the compounds.}
\label{fig:fig3}
\end{figure}

The crystal structure of Y$_3$Ru$_4$Ge$_{13}$ is shown in Fig.~\ref{fig:fig1}. Both compounds have same crystal structure and cubic symmetry ($\it{Pm3n}$, space group $\#$ 223). Rietveld analysis \cite{Carvajal1993} of the Powder X-ray diffraction (PXRD) of Y$_3$Ru$_4$Ge$_{13}$ is shown in  Fig.~\ref{fig:fig2}. The temperature dependence of resistivity $\rho (T)$ from 300 to 2K for Y$_3$Ru$_4$Ge$_{13}$ and Lu$_3$Os$_4$Ge$_{13}$  are shown in Fig.~\ref{fig:fig3}. A semi-metallic behaviour $( \frac{d\rho}{dT}<0 )$ can be observed in the normal state resistivity data of both the compounds. 
\begin{figure}[h]
\hfill
\subfigure{\includegraphics[width=7cm]{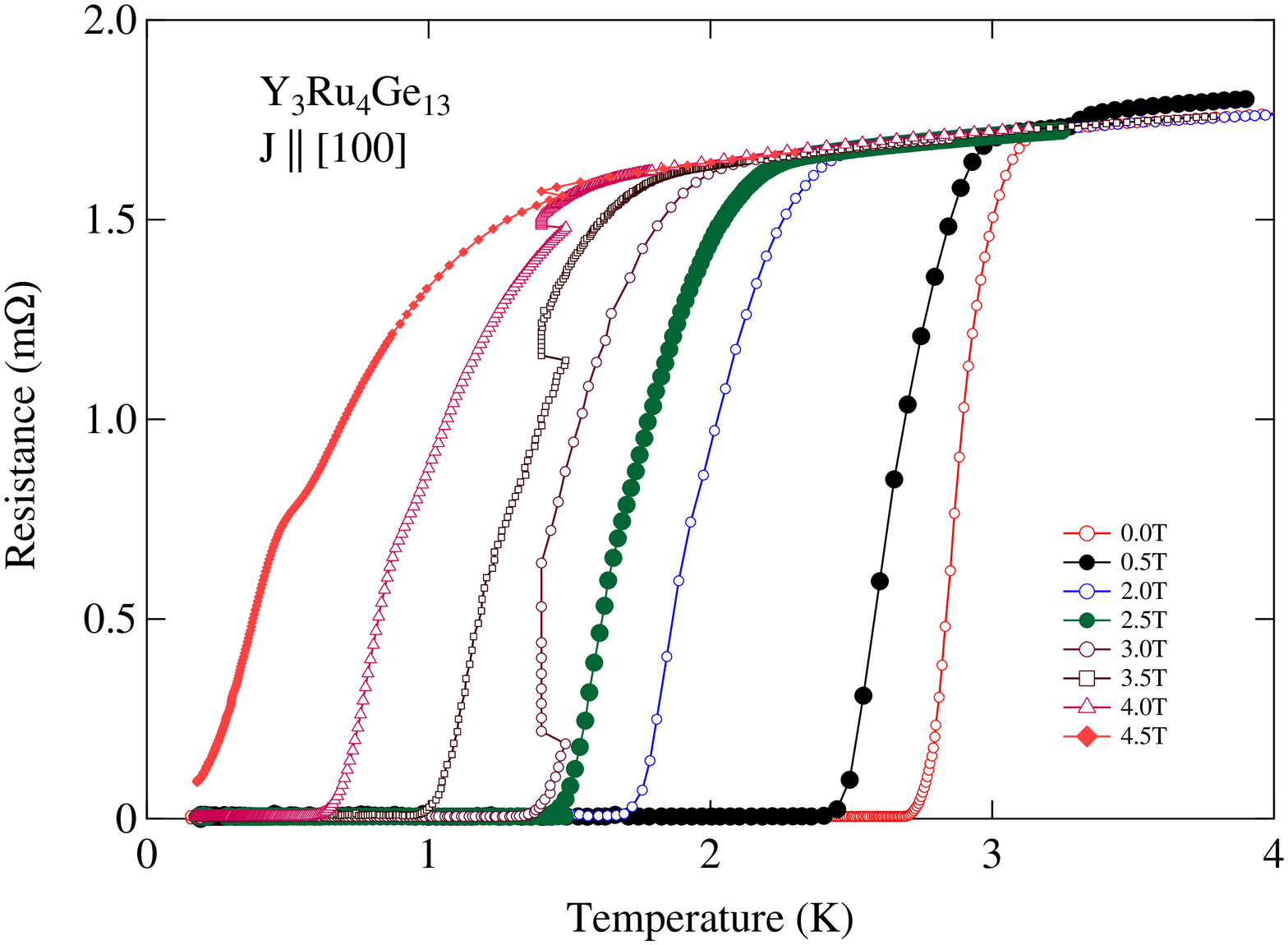}}
\label{fig:fig3a}
\hfill
\subfigure{\includegraphics[width=7cm]{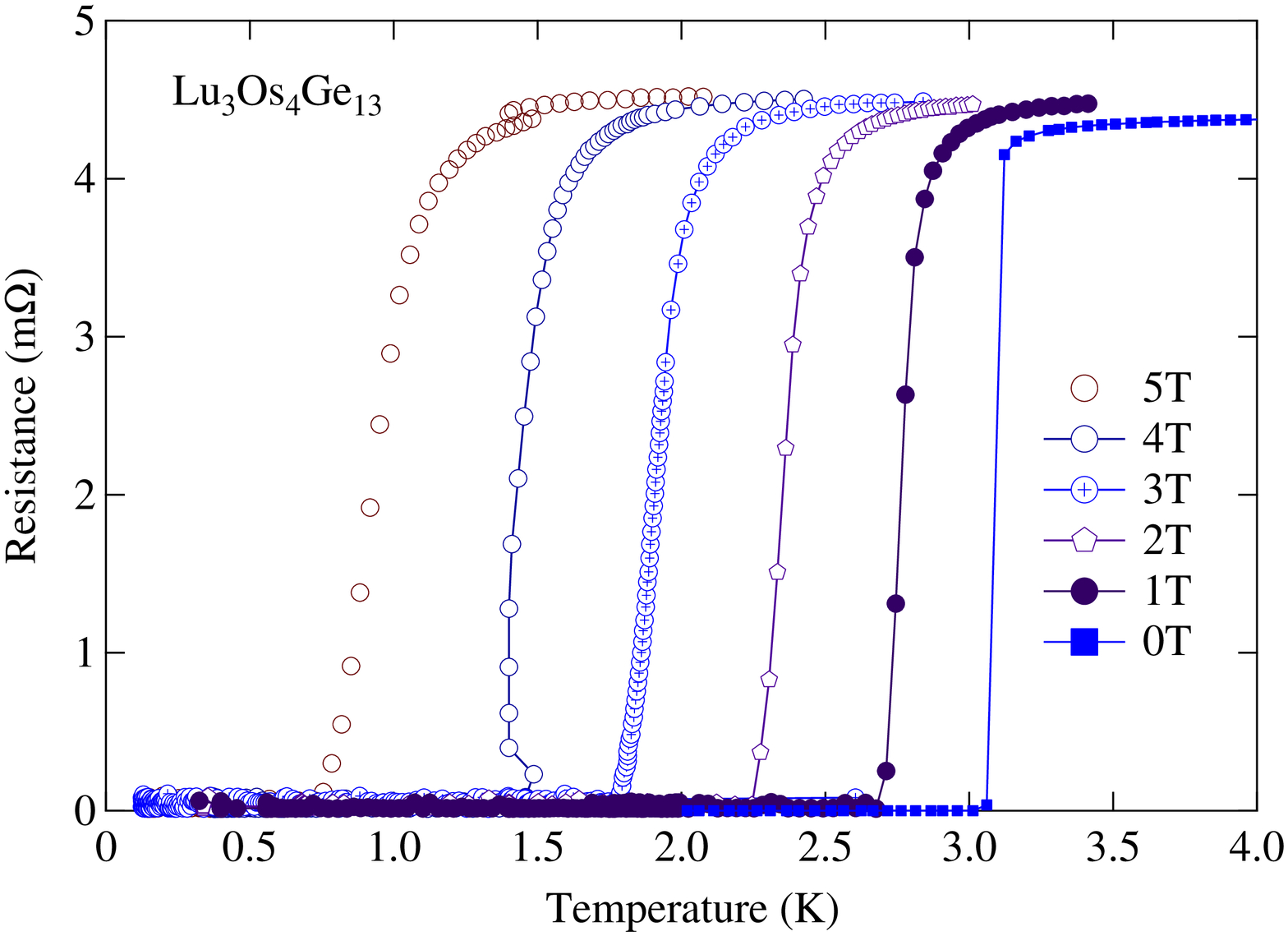}}
\label{fig:fig3b}
\hfill
\caption{The resistance vs temperature data at different magnetic fields for Y$_3$Ru$_4$Ge$_{13}$ and Lu$_3$Os$_4$Ge$_{13}$. With the increase in applied magnetic field the superconducting transition becomes slightly broader for both compounds. The temperature dependence of the upper critical field ($\mu{_0}Hc{_2}(T)$) is extracted from these measurements.}
\label{fig:fig4}
\end{figure}
The magnetoresistance data for Y$_3$Ru$_4$Ge$_{13}$ and Lu$_3$Os$_4$Ge$_{13}$ is shown in Fig.~\ref{fig:fig4}. The width of the superconducting transition increases with increasing magnetic field. The transition temperature is taken at the point where resistivity becomes half of its normal state value. 
The temperature dependence of the upper critical field for Y$_3$Ru$_4$Ge$_{13}$ and Lu$_3$Os$_4$Ge$_{13}$ is shown in Fig.~\ref{fig:fig5}. We estimate the orbital upper critical field, $\mu{_0}Hc{_2}(0)$, for both the compounds using Werthamer-Helfand-Hohenberg (WHH) expression\cite{Werthamer1966}, $\mu{_0}Hc{_2}(0) = -0.693 ~T{_c}\frac{dHc{_2}}{dT}\vert{_{T=T{_c}}}$ in the dirty limit for type-II superconductors. A nearly linear relationship is observed in Fig.~\ref{fig:fig5} between $\mu{_0}Hc{_2}$ and $T{_c}$ in the proximity of the transition temperature ($T{_c}$ at H = 0) for both the compounds but the linear trend is more prominent for Lu$_3$Os$_4$Ge$_{13}$. 
\begin{figure}[ht]
\hfill
\subfigure{\includegraphics[width=7cm]{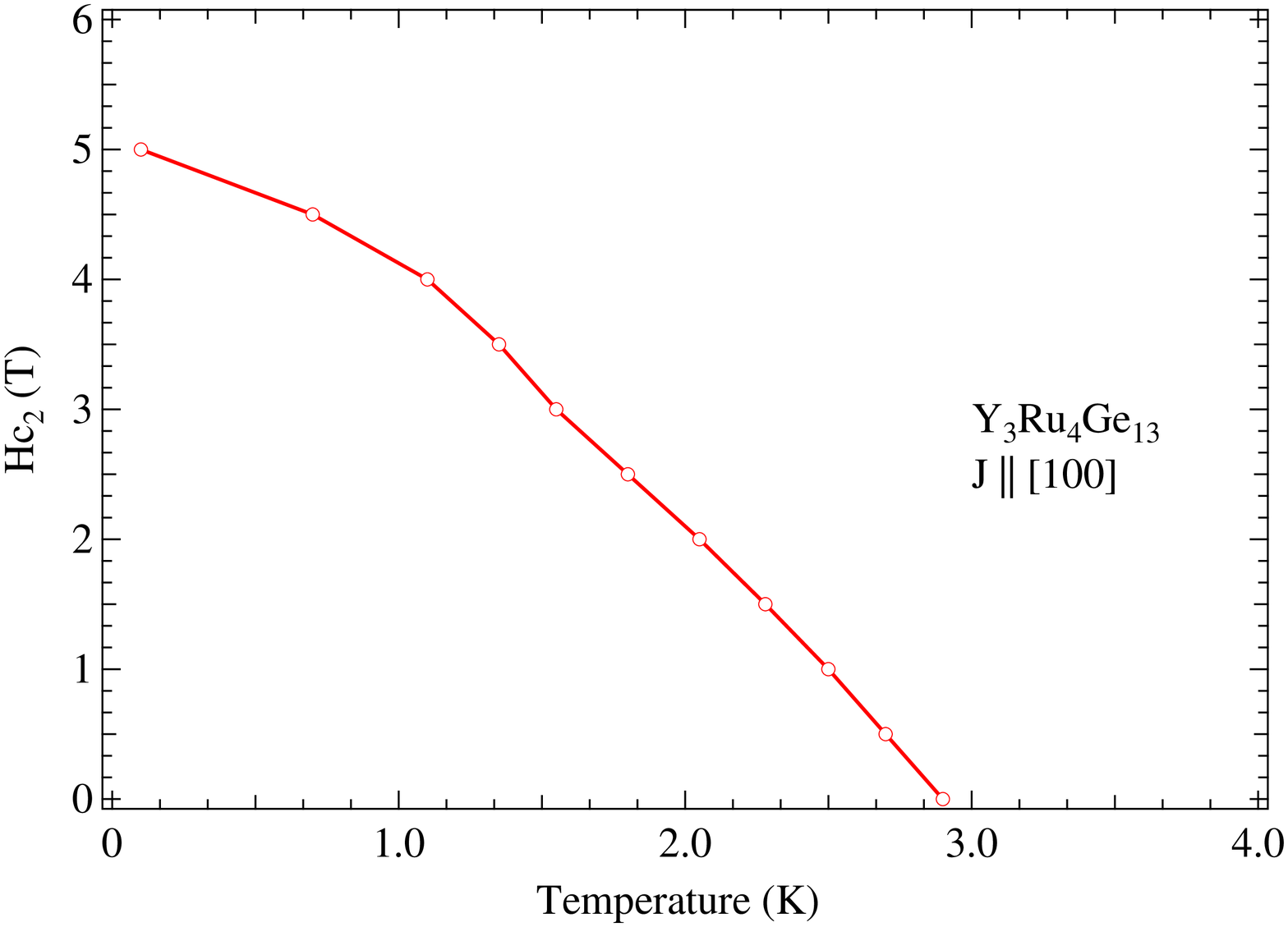}}
\label{fig:fig4a}
\hfill
\subfigure{\includegraphics[width=7cm]{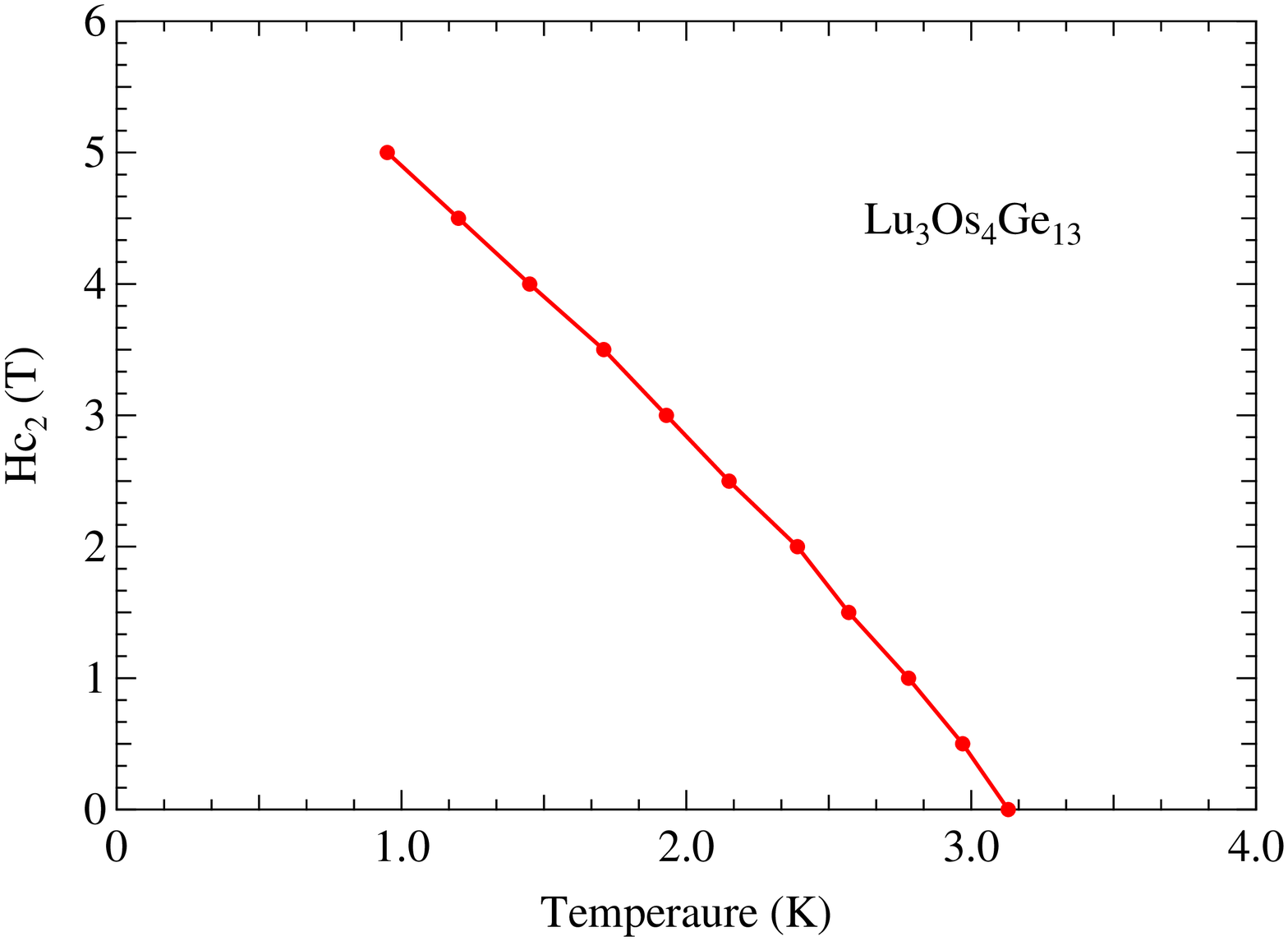}}
\label{fig:fig4a}
\hfill
\caption{$\mu{_0}H{_{c2}}$ as a function of temperature for Y$_3$Ru$_4$Ge$_{13}$ and Lu$_3$Os$_4$Ge$_{13}$. The upper critical field  $\mu{_0}H{_{c2}}$  increases linearly as the temperature is lowered in the vicinity of the transition temperature $T_{c}$ for both compounds, though the linear dependence is more prominent for Lu$_3$Os$_4$Ge$_{13}$.}
\label{fig:fig5}
\end{figure}
The slope $\frac{dHc{_2}}{dT}\vert{_{T=T{_c}}}$ is used to calculate $\mu{_0}H{_{c2}} = 4.63\pm0.09$~T for Y$_3$Ru$_4$Ge$_{13}$ and $\mu{_0}Hc{_2}= 5.68\pm 0.12$~T for Lu$_3$Os$_4$Ge$_{13}$ using the WHH formula in the dirty limit. The value of $\mu{_0}H{_{c2}}$ is smaller than the weak coupling Pauli paramagnetic limit $\mu{_0}H{^\mathrm{Pauli}}=1.82T_{c}=5.09$~T for Y$_3$Ru$_4$Ge$_{13}$ and $\mu{_0}H{^\mathrm{Pauli}}=5.80$~T for Lu$_3$Os$_4$Ge$_{13}$. The upper critical field value $\mu{_0}H{_{c2}}(0)$ can be used to estimate the Ginzburg-Landau coherence length $\xi(0){_{GL}}=\sqrt{\Phi{_0}/{2\pi{H{_{c2}}(0)}}}= 80.4\pm{0.5}\AA$~for Y$_3$Ru$_4$Ge$_{13}$ and $\xi(0){_{GL}}= 76.1\pm{0.7}\AA$~for Lu$_3$Os$_4$Ge$_{13}$, where $\Phi{_0}={hc}/{2e}$ is the magnetic flux quantum.\\
The DC-magnetisation data of both compounds is shown in Fig.~\ref{fig:fig6} indicating diamagnetic transitions of Y$_3$Ru$_4$Ge$_{13}$ at 2.8~K and Lu$_3$Os$_4$Ge$_{13}$ at 3.1~K. Very similar values of  $T_{c}$ from both resistivity and susceptibility data confirm that our single crystals are of very high quality. Large vortex pinning can be observed in the field cooled (FC-Meissner) data in shown in Fig.~\ref{fig:fig6}.
\begin{figure}[h]
\hfill
\subfigure{\includegraphics[width=7cm]{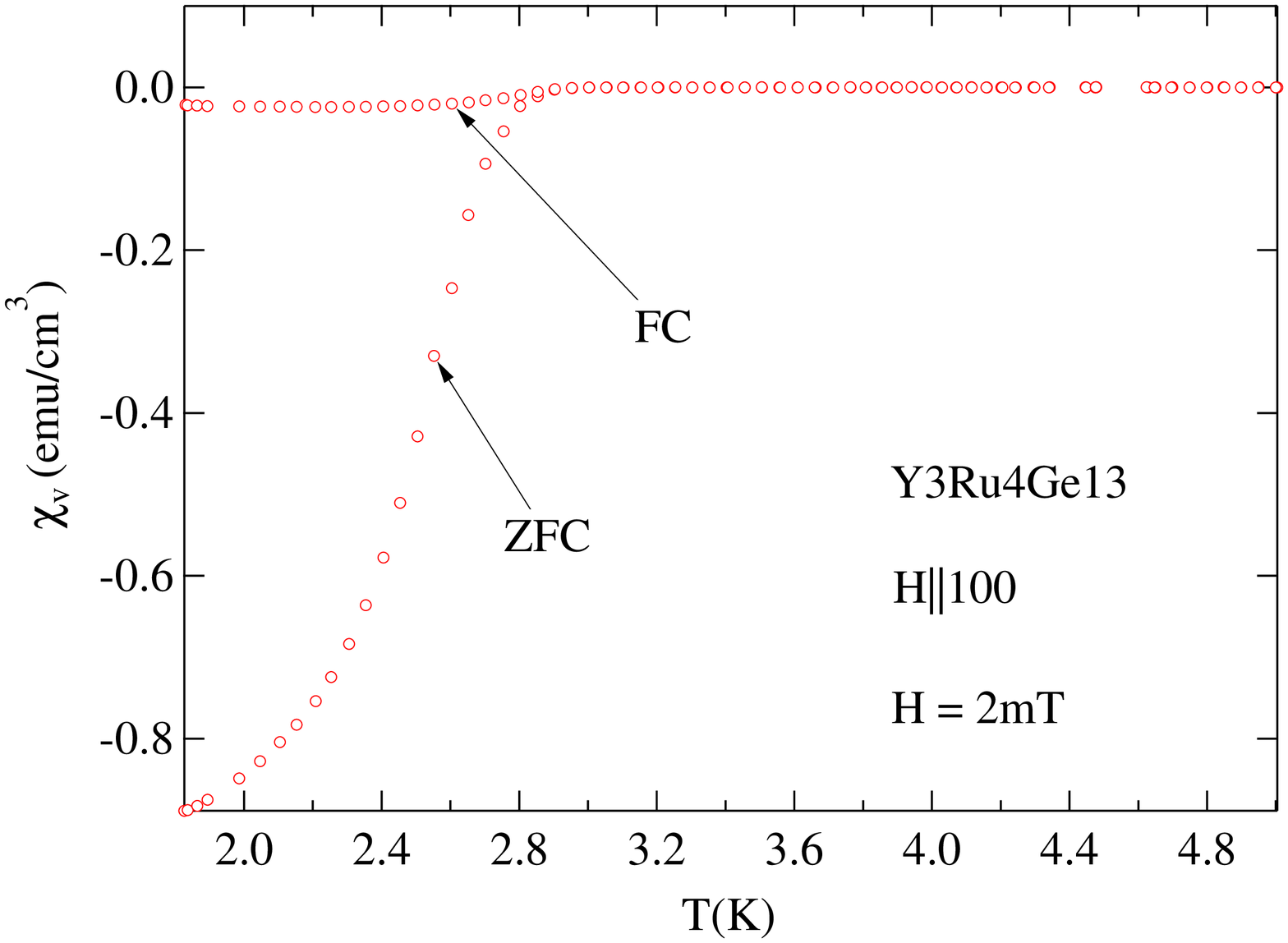}}
\label{fig:fig5a}
\hfill
\subfigure{\includegraphics[width=7cm]{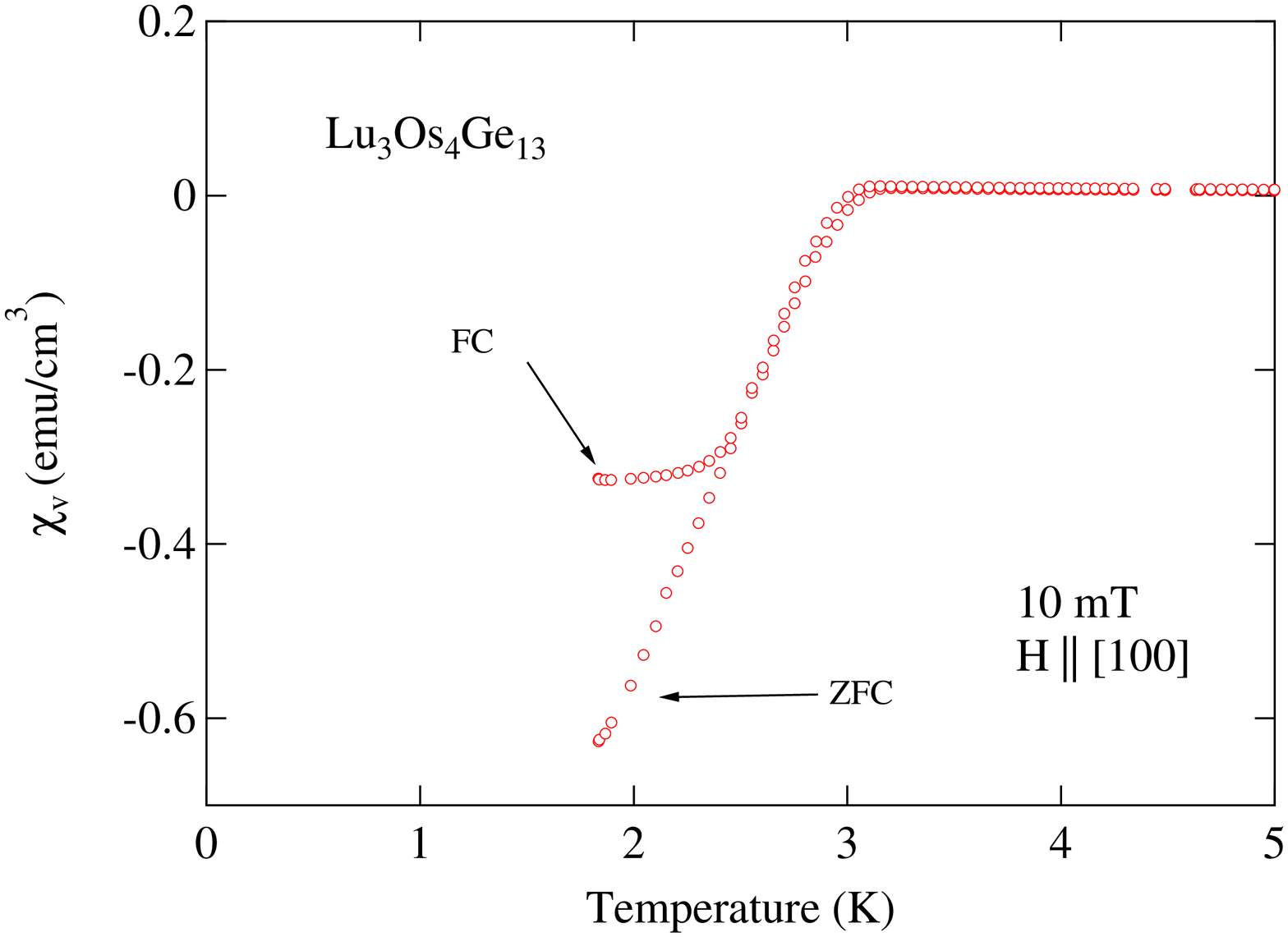}}
\label{fig:fig5b}
\hfill
\caption{DC magnetic susceptibility as a function of temperature for magnetic field $H\parallel[100]$ direction for Y$_3$Ru$_4$Ge$_{13}$ and Lu$_3$Os$_4$Ge$_{13}$. The superconducting transition temperatures determined from susceptibility measurements are in excellent agreement with the resistivity data reflecting the high quality of the single crystals. The ZFC and FC susceptibility data indicate significant amount of pinning of vortices in both the compounds.}
\label{fig:fig6}
\end{figure}
\begin{figure}[ht]
\hfill
\subfigure{\includegraphics[width=7cm]{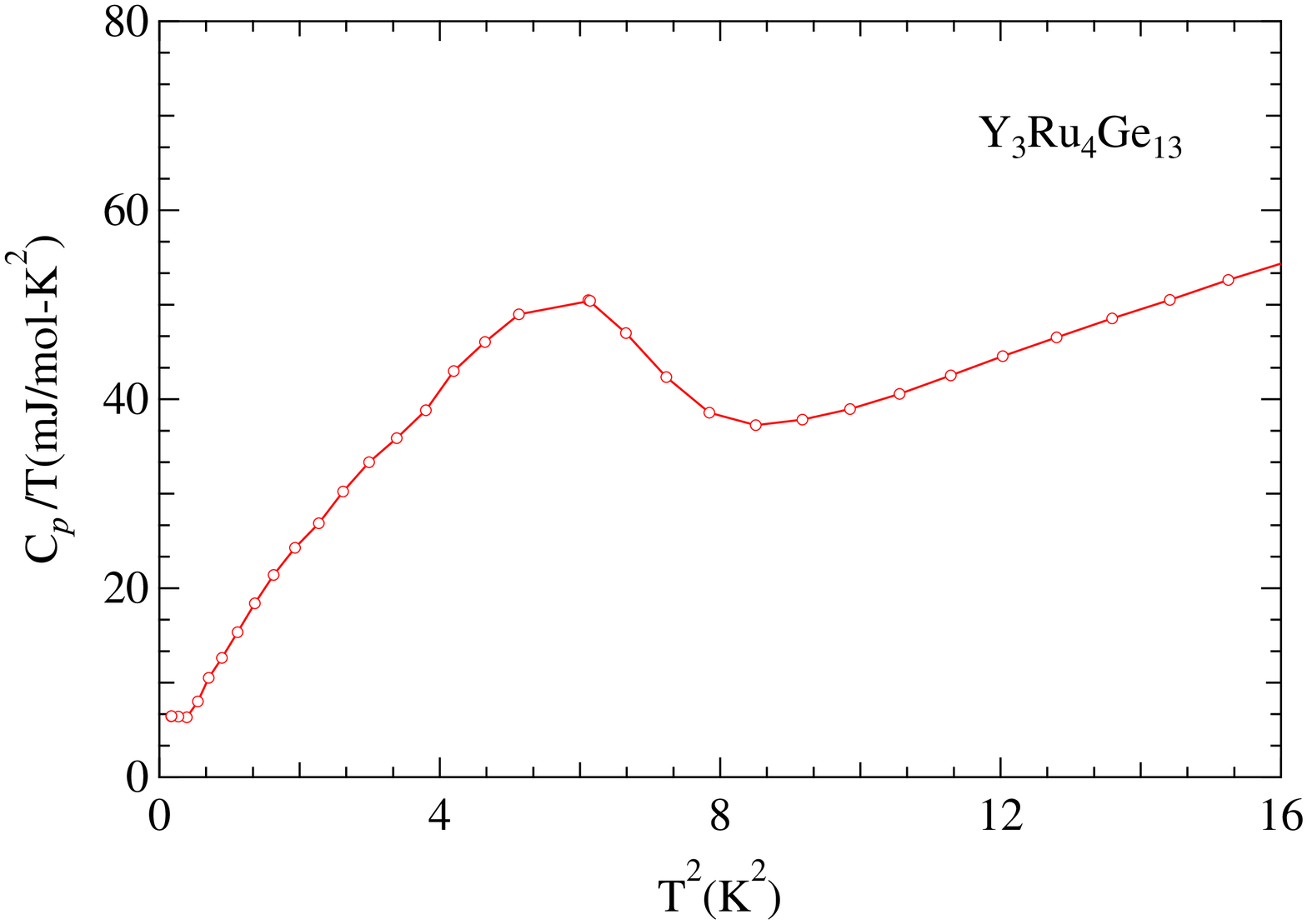}}
\label{fig:fig6a}
\hfill
\subfigure{\includegraphics[width=7cm]{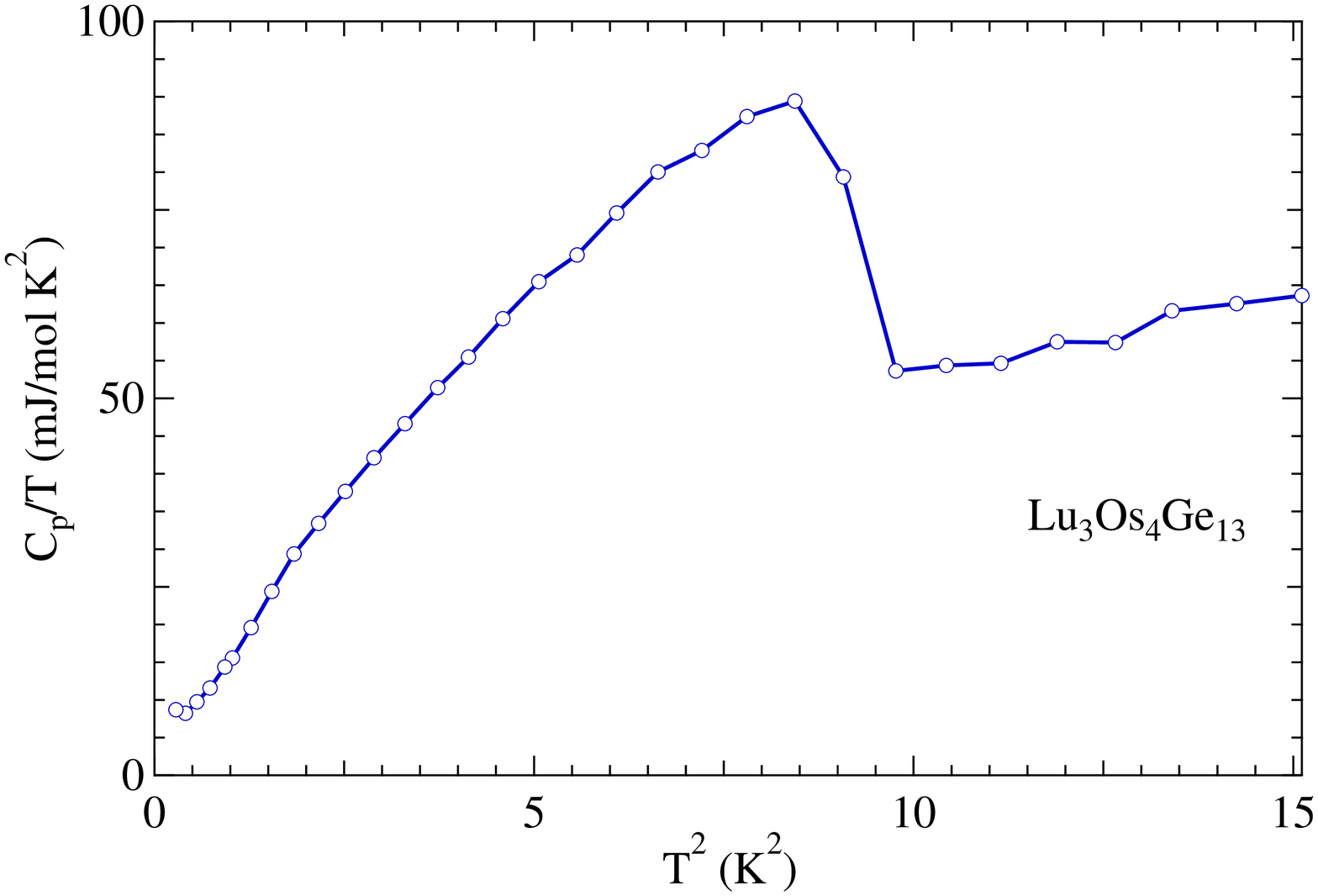}}
\label{fig:fig6b}
\hfill
\caption{{C/T} vs $T{^2}$ data for Y$_3$Ru$_4$Ge$_{13}$ and Lu$_3$Os$_4$Ge$_{13}$ respectively. Large jump in heat capacity confirms bulk superconductivity in both the compounds.}
\label{fig:fig7}
\end{figure}
The characterisation of the superconducting transition using heat capacity measurements is shown in Fig.~\ref{fig:fig7}.  The specific heat jump at the thermodynamic transition confirm the bulk superconductivity in both the compounds. The low temperature normal state specific heat can be well fitted with $\frac{C}{T} = \gamma + \beta T^2$, where $\gamma T$ represents the electronic contribution and $\beta T^3$ describe the lattice-phonon contributions to the specific heat in the normal state. Fitting the above formula give electronic specific heat coefficient $\gamma= 7.08\frac{mJ}{mol K^2}$ ($\gamma =25.4 \frac{mJ}{mol K^2}$ ) and the phonon/lattice contributions $\beta = 3.52\frac{mJ}{mol K^4}$$(\beta = 2.30\frac{mJ}{mol K^2})$ for Y$_3$Ru$_4$Ge$_{13}$ (Lu$_3$Os$_4$Ge$_{13}$). The ratio $\frac{\Delta C}{\gamma Tc}$ can be used to measure the strength of the electron-phonon coupling. The specific heat jump $\frac{\Delta C}{Tc}$ is $6.07\frac{mJ}{mol K^2}$  ($29\frac{mJ}{mol K^2}$), setting the value of  $\frac{\Delta C}{\gamma Tc} = 0.85$ ( $\frac{\Delta C}{\gamma Tc} = 1.15$ ) for Y$_3$Ru$_4$Ge$_{13}$ (Lu$_3$Os$_4$Ge$_{13}$). These values are smaller than the weak-coupling limit of 1.43 for a conventional BCS superconductor, suggesting that these two compounds are moderately electron-phonon coupled superconductor.\\
The comparison among the normal and superconducting state parameters of the both compounds is shown in Table-\ref{table:table1}. We also notice from Table-\ref{table:table1} that value of  $\gamma$ is larger in Lu$_3$Os$_4$Ge$_{13}$ suggesting stronger electronic correlations in Lu$_3$Os$_4$Ge$_{13}$ as compared to electronic correlations in Y$_3$Ru$_4$Ge$_{13}$. 
\begin{center}
\begin{table}[h]
\centering
\caption{\label{tab:params}  Normal and superconducting state parameters of Y$_3$Ru$_4$Ge$_{13}$ and Lu$_3$Os$_4$Ge$_{13}$}
\begin{tabular}{lcc}
\br
Parameters			&Y$_3$Ru$_4$Ge$_{13}$ & Lu$_3$Os$_4$Ge$_{13}$\\
\mr
$T_{\rm c}$~(K)                                &2.85         &  3.1   \\
$\gamma$~(mJ/mol\,K$^{2}$)            &7.1          & 25.4 \\
$\Theta_{\rm D}$~(K)                            &223        & 257 \\
$\Delta C_{\rm el}/\gamma T_{\rm c}$          &0.85   & 1.15 \\
$\mu{_0}H{_{c2}}$~(T)			&4.63		&5.68\\
$\xi(0)_{GL}$ ($\AA$)                         &80.4                &  78 \\
\br
\end{tabular}
\label{table:table1}
\end{table}
\end{center}
\section{Conclusion}
We have grown single crystals and characterised the superconducting properties of two semi-metallic compounds Y$_3$Ru$_4$Ge$_{13}$ and Lu$_3$Os$_4$Ge$_{13}$. A bulk superconducting transition is confirmed and characterised through electrical transport, magnetisation and heat capacity measurements on the single crystals. The magnetic susceptibility measurements show large pinning of vortices in both the compounds. The analysis of the low temperature heat capacity data suggests that both these compounds are moderately electron-phonon coupled type-$II$ superconductors.
\section*{References}
\raggedright
\bibliography{reference}
\end{document}